\documentclass[aps,pre,twocolumn,titlepage,nofootinbib]{revtex4}

\usepackage{graphicx}
\usepackage{color}
\usepackage{dcolumn}
\usepackage{latexsym}
\usepackage[normalem]{ulem}
\usepackage{hyperref,amssymb}
\usepackage{url}
\usepackage{color,soul}
\usepackage{graphicx}
\usepackage{verbatim}
\usepackage{multirow}
\usepackage{amsmath}
\newcommand{\beq}{\begin{eqnarray}}
\newcommand{\eeq}{\end{eqnarray}}
\usepackage{mathrsfs}
\usepackage{float,soul}
\usepackage[dvipsnames]{xcolor}
\usepackage{mathtools}
\usepackage{slashed}
\usepackage{physics}	
\usepackage{graphicx}   
\usepackage{epstopdf}
\usepackage{subfigure}  
\usepackage{hyperref}   
\usepackage{bbold}
\usepackage{wasysym}
\usepackage{feynmp}
\usepackage{hyperref}
\hypersetup{colorlinks,
}

\usepackage[latin1]{inputenc}

\begin{document}

\title{\Large Glassy heat capacity from overdamped phasons and a hypothetical phason-induced superconductivity in incommensurate structures}
\author{Cunyuan Jiang$^{1,2,3}$}
\author{Alessio Zaccone$^4$}
\author{Chandan Setty$^5$}
\author{Matteo Baggioli$^{1,2,3}$}
\thanks{b.matteo@sjtu.edu.cn}
\address{$^1$ School of Physics and Astronomy, Shanghai Jiao Tong University, Shanghai 200240, China}
\address{$^2$ Wilczek Quantum Center, School of Physics and Astronomy, Shanghai Jiao Tong University, Shanghai 200240, China}
\address{$^3$ Shanghai Research Center for Quantum Sciences, Shanghai 201315,China}
\address{$^4$ Department of Physics ``A. Pontremoli'', University of Milan, via Celoria 16,
20133 Milan, Italy}
\address{$^5$ Department of Physics and Astronomy, Rice Center for Quantum Materials, Rice University, Houston, Texas 77005, USA}

\begin{abstract}
Phasons are collective low-energy modes that appear in disparate condensed matter systems such as quasicrystals, incommensurate structures, fluctuating charge density waves, and Moir\'e superlattices. They share several similarities with acoustic phonon modes, but they are not protected by any exact translational symmetry. As a consequence, they are subject to a wavevector independent damping, and they develop a finite pinning frequency, which destroy their acoustic linearly propagating dispersion. Under a few and simple well-motivated assumptions, we compute the phason density of states, and we derive the phason heat capacity as a function of the temperature. Finally, imagining a hypothetical s-wave pairing channel with electrons, we compute the critical temperature $T_c$ of the corresponding superconducting state as a function of phason damping using the Eliashberg formalism. We find that for large phason damping, the heat capacity is linear in temperature, showing a distinctive glass-like behavior. Additionally, we observe that the phason damping can strongly enhance the effective Eliashberg coupling, and we reveal a sharp non-monotonic dependence of the superconducting temperature $T_c$ on the phason damping, with a maximum located at the underdamped to overdamped crossover scale. Our simple computations confirm the potential role of overdamped modes in explaining the glassy properties of incommensurate structures, but also in possibly inducing strongly-coupled superconductivity therein, and enhancing the corresponding $T_c$.
\end{abstract}

\maketitle

\section{Introduction}
Acoustic phonons are collective vibrational modes which appear in crystalline solids because of the spontaneous breaking of translations (long-range order) \cite{Lubensky}. As Goldstone modes \cite{leutwyler1996phonons}, their dispersion relation is protected by symmetries, hence, phonons\footnote{In this work, we will not consider optical modes. Therefore, when we mention phonons, we will always refer to acoustic ones.} are always gapless modes. Their dispersion relation can be obtained by solving the following equation
\begin{equation}\label{d1}
    \omega^2=c^2 k^2 -i \omega D k^2+\dots,
\end{equation}
where $\omega, k$ are respectively the frequency and wave-vector and the $\dots$ indicated higher-order terms. Here, for simplicity, we have assumed isotropy and neglected any distinction between the different branches (longitudinal or transverse). In the above equation, $c$ and $D$ are the phonon speed and the phonon attenuation constant. Eq.\eqref{d1} can be derived using several methods, see for example \cite{Lubensky,RevModPhys.95.011001}.

Many of the low-energy vibrational and thermodynamic properties of solids (\textit{e.g.}, density of states, heat capacity, etc.) can be efficiently rationalized starting from the simple concept of phonons, within the celebrated Debye's theory \cite{kittel2021introduction}. In metallic systems, further properties, such as the electric resistivity or the onset of superconductivity, are determined, or at least strongly affected, by phonons and their coupling to electrons. Typical cases are given by the dominant phonon-electron scattering at high-temperature \cite{ziman2001electrons}, which gives a linear in $T$ resistivity above the Debye's temperature, or the phonon-electron coupling which is behind the Bardeen-Cooper-Schrieffer (BCS) theory of superconductivity \cite{Carbotte2003}.

Aperiodic crystals are complex structures which do not display full translational order as in periodic crystalline structures, but still enjoy long-range order, hence they are solid. Typical examples of that sort are quasicrystals, incommensurate structures, fluctuating charge density waves, and Moir\'e superlattices. In all these different scenarios, in addition to the standard acoustic phonons, extra low-energy modes appear (\textit{e.g.}, \cite{PhysRevB.106.054113,PhysRevLett.99.035501,doi:10.1080/00150198608227873,ZEYEN1983283,QUILICHINI19831011,refId0}). Depending on the context, these modes are usually referred to as sliding modes, phasons, or Moir\'e phonons. For simplicity, we will refer to them with the collective label ``phason''. In composite crystals, or incommensurate charge density wave systems, these modes correspond to the relative rigid translations of the two superstructures along the incommensurate direction. In modulate structures, the phason, as the name indicates, it is just the fluctuation of the phase of the static modulation wave. In quasicrystals, the phason corresponds to translations in the direction perpendicular to the irrational cut in the extra-dimensional superspace picture \cite{DeBoissieu2008-DEBPMI-3,https://doi.org/10.1002/ijch.201100131}, and it is harder to visualize (and indeed its meaning is still controversial \cite{doi:10.1080/14786430802247163}). See \cite{CURRAT1988201} for a nice review about excitations in incommensurate phases of various types. Notice that, in many situations (for example in quasicrystals), phasons do not correspond to the translations of atoms but rather to particles rearrangements, more \textit{akin} to the diffusive motion in liquids (see for example \cite{PhysRevLett.108.218301}).

\begin{figure}
    \centering
    \includegraphics[width=0.8\linewidth]{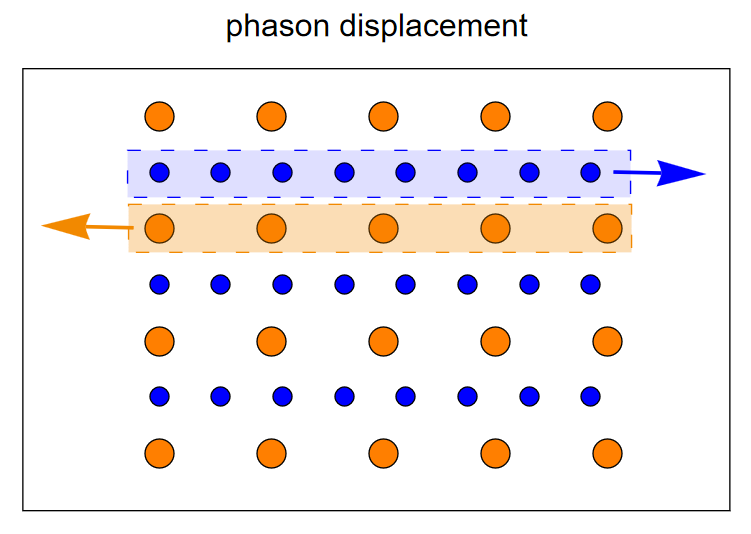}
    \caption{A visualization of the phason displacement in an incommensurate structure with two incommensurate lattices. A concrete example is that of the mercury chain compound Hg$_{3-\delta}$AsF$_6$ \cite{PhysRevB.18.3645,PhysRevLett.39.1484}. The phason corresponds to a relative rigid translation between the two lattices.}
    \label{fig0}
\end{figure}

Importantly, no matter the specific microscopic structure from which they arise, phasons do not correspond to an exact translational symmetry and therefore, differently from acoustic phonons, they are not protected by symmetry. In mathematical terms, this implies that their dispersion relation follows from an equation of the form
\begin{equation}
    \omega^2-\Omega(k)^2+i \omega \Gamma(k) =0\,,
\end{equation}
where $\Omega(k),\Gamma(k)$ are not constrained to vanish in the limit of $k \rightarrow 0$. In other words, the phason dispersion relation can be characterized by a pinning frequency $\omega_0$ and a friction/damping term $\gamma$, given by:
\begin{equation}
    \Gamma(0)\equiv \gamma\,,\qquad \Omega(0)\equiv \omega_0\,.
\end{equation}
Both terms are not allowed for acoustic phonons. For a discussion about the pinning frequency in charge density waves, see for example \cite{RevModPhys.60.1129}. Interestingly, for pseudo-Goldstone modes, the two parameters above, $\omega_0,\gamma$, are indeed related \cite{RevModPhys.95.011001} in a way similar to the famous Gell-Mann-Oakes-Renner relation for Pions.

In the limit of weak quenched disorder, or equivalently large anharmonic interactions and dissipation, the pinning frequency term can be neglected, since $\omega_0\ll \gamma$ (see \cite{PhysRevLett.128.065901} for an estimate in twisted bilayer graphene). Hence, the dispersion relation for the phason can be well approximated by solving the following equation:
\begin{equation}
    \omega^2+i \omega \gamma=v^2 k^2\,,
\end{equation}
where $v,\gamma$ are respectively the phason speed and friction (see the next section for more details). This structure can be derived in several ways, from hydrodynamics \cite{PhysRevB.32.7444} and vibrational models \cite{Currat2002}, to arguments based purely on symmetries and effective field theory \cite{10.21468/SciPostPhys.9.5.062}. Interestingly, the dispersion which arises from solving the above equation is identical to that expected for shear waves in liquids \cite{BAGGIOLI20201}, and in general for systems described by quasi-hydrodynamics \cite{PhysRevD.99.086012}. In particular, there appears a crossover between a overdamped regime to an underdamped propagating one by decreasing the wavelength \cite{PhysRevLett.49.468,PhysRevLett.49.1833,PhysRevB.28.340,Currat2002,PhysRevB.26.4963}. Both behaviors have been experimentally observed respectively using inelastic neutron-scattering experiments in biphenyl \cite{CAILLEAU1980407}, and light-scattering experiments in BaMnF$_4$ \cite{PhysRevB.25.1791}.

Aside from their peculiar dispersion, phason modes have been identified as the fundamental origin for the glassy-like properties of incommensurate structures and aperiodic crystals, which have been revealed in several instances in the literature \cite{PhysRevB.99.054305,PhysRevLett.93.245902,PhysRevLett.114.195502}. More precisely, the phason gap, $\omega_0$, has been shown to give an excess in the heat-capacity \textit{akin} to that observed in glasses, \textit{a.k.a.}, the boson peak \cite{doi:10.1142/q0371}. The role of the pinned phason is similar in spirit to the possible contributions from soft optical modes \cite{Schliesser_2015,Baggioli_2020}, which are evident in thermoelectrics \cite{RevModPhys.86.669} and organic compounds \cite{PhysRevB.99.024301,doi:10.1021/acs.jpclett.2c01224}. There is clear evidence of the role of the phason gap in this regard \cite{PhysRevLett.114.195502,PhysRevLett.93.245902,PhysRevLett.76.2334}. 

On top of that, the overdamped nature of phasons for low wavevectors has been shown to produce a linear in $T$ contribution to the heat capacity, similar to that of two-level states (TLS) in glasses \cite{PhysRevLett.114.195502,PhysRevLett.93.245902,PhysRevLett.76.2334,PhysRevResearch.1.012010,doi:10.1142/S0217979221300024}. A quasi-linear scaling of the heat capacity has indeed been observed in several incommensurate compounds \cite{PhysRevB.34.4432,PhysRevLett.76.2334,Odin2001}. Let us also emphasize the similarity between phasons in incommensurate structures and vortons in superconducting vortex lattices.\footnote{We thank Hector Ochoa for pointing this out to us.} Both excitations are acoustic-like bosonic modes which appear overdamped for large wavelenghts. Not coincidentally, a linear in $T$ heat capacity in superconducting vortex lattices has been already found, and attributed to the vorton contribution \cite{PhysRevLett.71.3541,PhysRevB.50.10272,PhysRevB.57.6046}. We expect the features described in this work, and related literature, to be generic in presence of overdamped acoustic-like modes. Interestingly, further relations between the physics of incommensurate orders (and specially pinned fluctuating charge density waves) and that of glasses have been discovered in the past \cite{PhysRevB.41.11522,PhysRevB.38.2675}.

A less investigated feature of phasons is their coupling to electrons, their role in electronic transport, and finally in possible superconducting instabilities. This topic has been recently revived by the growing interest around twisted bilayer graphene (TBG), in which a phason mode is expected as a result of the rotational degree of freedom between the two layers \cite{PhysRevB.100.075416,PhysRevResearch.2.013335,PhysRevB.106.075420,PhysRevLett.128.065901}. The role and strength of electron-phason coupling have been recently investigated in \cite{PhysRevB.100.155426,PhysRevLett.128.065901}, and the effects of phason-electron scattering in \cite{ochoa2023extended}.

Given its overdamped nature, the phason is expected to originate additional spectral weight at very low energy which can potentially enhance the effective coupling to electrons and therefore superconductivity. This effect might appear as general in incommensurate structures, and it has been already reported experimentally in a quasiperiodic host-guest structure of elemental bismuth at high pressure, Bi-III \cite{doi:10.1126/sciadv.aao4793}. The superconductivity enhancement induced by incommensurability has also been recently advocated in the weak and intermediate coupling regimes in \cite{oliveira2023incommensurability}. The role of gapped phasons has been also investigated in incommensurate host-guest phases in compressed elemental sulfur in \cite{PhysRevB.103.214111}. More in general, the application of high-pressure can transform a single element metal into an incommensurate guest-host structure, as for the case of Scandium \cite{D1CP05803G}. Pressure can therefore be used as an external dialing parameter to adjust and control commensurability, and therefore the properties of the corresponding phason, providing a big potential to gauge superconductivity. Finally, phasons could play a fundamental role in the recently discovered superconductivity of quasicrystals \cite{Kamiya2018}, but also in cuprate oxide
high-Tc superconductors, where fluctuating charge density wave order exists in the vicinity of the superconducting phase (\textit{e.g.}, \cite{Torchinsky2013,PhysRevLett.96.137003,arpaia2022signature}), and it might be responsible for several of the mysterious properties therein \cite{sym8060045,grilli2022disorder,caprara2022dissipation,seibold2021strange,arpaia2019dynamical}.

All in all, the motivations to study the effects of phasons on the mechanical, thermodynamic, transport and superconducting properties of incommensurate systems are many and timely. Following the Occam's razor principle, we present a concrete analysis in this direction based on few but simple assumptions. First, we take the limit in which the phason pinning frequency is subleading with respect to the phason damping $\gamma$. Then, we assume a s-wave isotropic pairing channel between phasons and electrons, as for the case of phonons.

In our work, we confirm the role of overdamped propagating modes for the glass-like properties of incommensurate structures at low temperature. Most importantly, we show using the Eliashberg formalism that overdamped phasons could lead to strongly coupled superconductivity and that their damping enhances the critical temperature $T_c$ in the underdamped case. Our findings are in agreement with the experimental results of \cite{doi:10.1126/sciadv.aao4793} and the recently theoretical analysis of \cite{oliveira2023incommensurability}. Also, the overdamped nature of the phason modes is ultimately due to anharmonicity, and it constitutes another interesting case for the sometimes fundamental role of the latter for superconductivity (see \cite{setty2023anharmonic} for a review).
\section{Phason modes}
We start by considering the phason Green's function given by
\begin{equation}\label{gr}
    \mathcal{G}(\omega,k)=\frac{1}{-\omega^2+v^2k^2-i \omega \gamma},
\end{equation}
where $\omega,k$ are respectively the phason frequency and wave-vector. Additionally, $v,\gamma$ are respectively the phason velocity and friction or damping parameter. The latter can be identified with the inverse relaxation time $\gamma \equiv \tau^{-1}$. Here, we have assumed that the pinning frequency $\omega_0$ of the phason, induced by disorder or impurities, is negligible. In other words, we assume $\gamma \gg \omega_0$. Notice that the phason dispersion relation obtained from Eq.\eqref{gr} is fundamentally different from the dispersion of acoustic and optical phonons. Acoustic phonons cannot display a wave-vector independent damping term since their dispersion is protected by translational symmetry. Also, optical phonons have a non-negligible $\omega_0$ which, at least in the most common case of underdamped optical modes, is much larger than the friction term $\gamma$. This can be partially understood by the fact that $\omega_0$ for phasons arises from explicitly breaking the emergent ``sliding'' symmetry (see for example \cite{10.21468/SciPostPhys.9.5.062}), while for optical phonons that is not the case as $\omega_0$ is simply a manifestation of non-universal microscopic physics. As a direct consequence, $\omega_0$ cannot be tuned to zero in the case of optical phonons, while it can in the case of phasons just assuming the absence of disorder and/or impurities. Following this assumption, the dispersion relation of the phason mode is given by
\begin{equation}\label{disp}
    \omega=-\frac{i\,\gamma}{2}\,\pm\,\sqrt{v^2k^2-\frac{\gamma^2}{4}}\,.
\end{equation}
\begin{figure}
    \centering
    \includegraphics[width=0.7\linewidth]{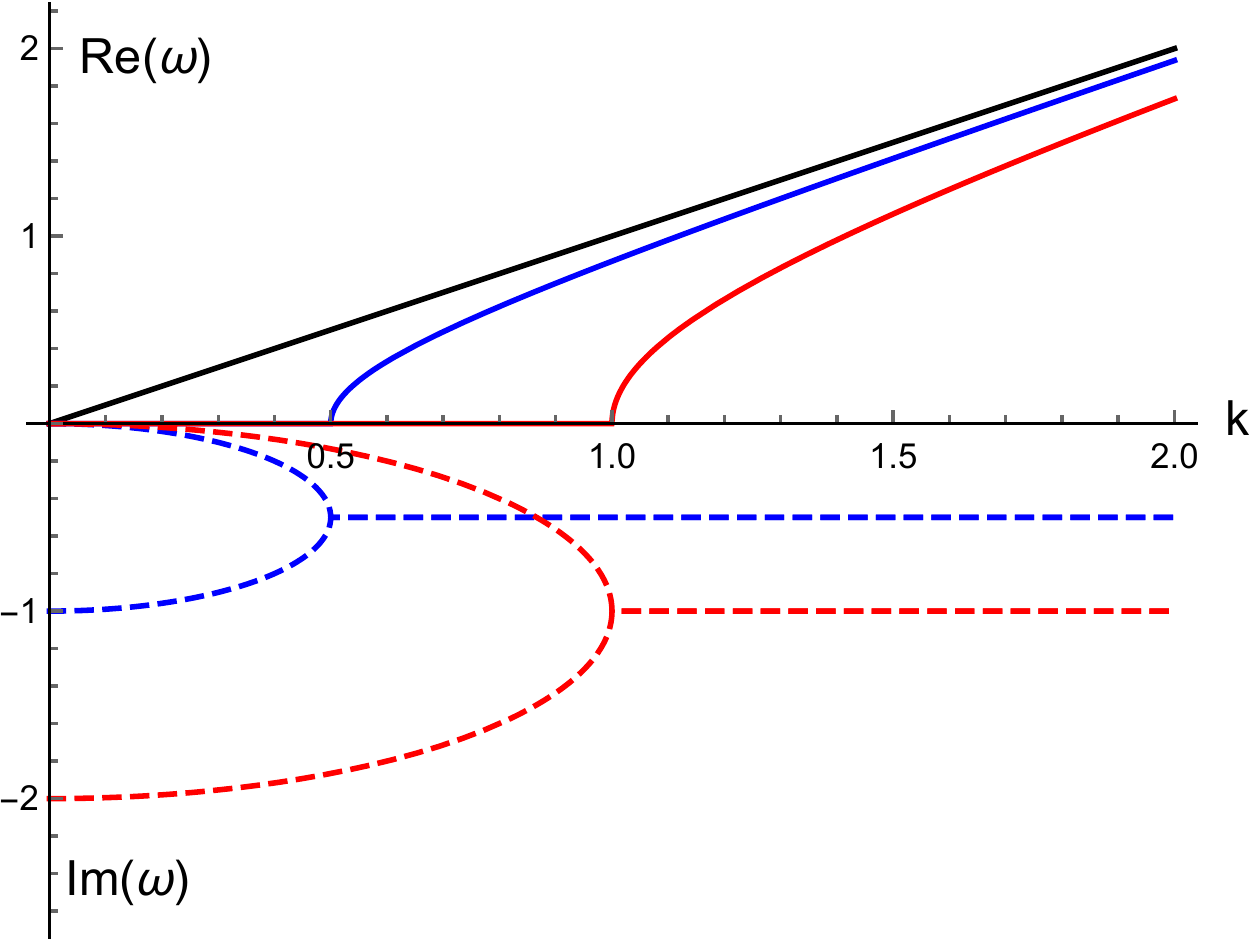}
    \caption{Phason dispersion relation upon increasing the friction parameter $\gamma$ and keeping the phason speed fixed. In this plot, $v=1$ and $\gamma=10^{-3},1,2$ from black to red. Filled and dashed lines are respectively the real and imaginary parts of the dispersion relation $\omega(k)$ in Eq.\eqref{disp}.}
    \label{fig1}
\end{figure}
The real and imaginary parts of the dispersion are shown in Fig.\ref{fig1}. In the long-wavelength limit, or equivalently for small wave-vectors, the phason is an overdamped mode with diffusive dispersion:
\begin{equation}\label{diff}
    \omega=-i D_\gamma k^2+\dots\quad D_\gamma=\,v^2/\gamma,
\end{equation}
and a second non-hydrodynamic damped mode, $\omega=-i \gamma$ appears as well.
We can then identify a crossover between the overdamped regime at low $k$ and an underdamped one at large $k$, in which the phason mode is propagating with a dispersion:
\begin{equation}
    \omega=\pm v k-\frac{i \gamma}{2}+\dots
\end{equation}
In such a regime, $\mathrm{Re}(\omega) \gg \mathrm{Im}(\omega)$. More precisely, by equating the real and imaginary parts of the dispersion, Eq.\eqref{disp}, we can find the crossover wave-vector, which turns out to be
\begin{equation}
    k^\star=\frac{\gamma}{\sqrt{2}v},
\end{equation}
or equivalently:
\begin{equation}
    \omega^\star=\frac{\gamma}{2}\,.
\end{equation}
Notice how this criterion, up to the $\sqrt{2}$ factor, reduces to the more familiar collisionless limit $\omega \tau \gg 1$.

Additionally, let us notice how the real part develops only after a certain critical wave-vector:
\begin{equation}
    \bar k=\frac{\gamma}{2 v}\,,
\end{equation}
which is smaller than the crossover scale. In other words, for $\bar k<k<k^\star$, the phason dispersion has a finite real part but the phason is still a non-propagating overdamped mode.

\begin{figure}
    \centering
    \includegraphics[width=0.7\linewidth]{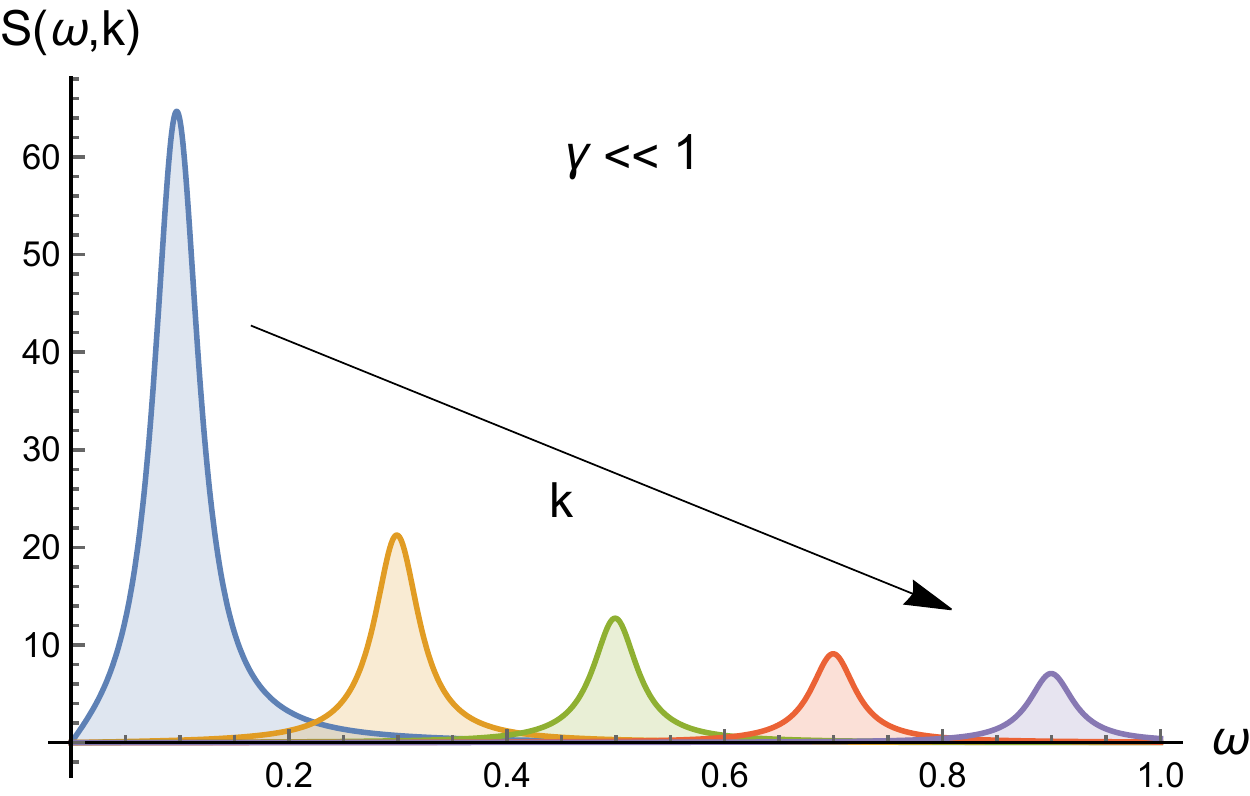}%

    \vspace{0.2cm}
    
    \includegraphics[width=0.7\linewidth]{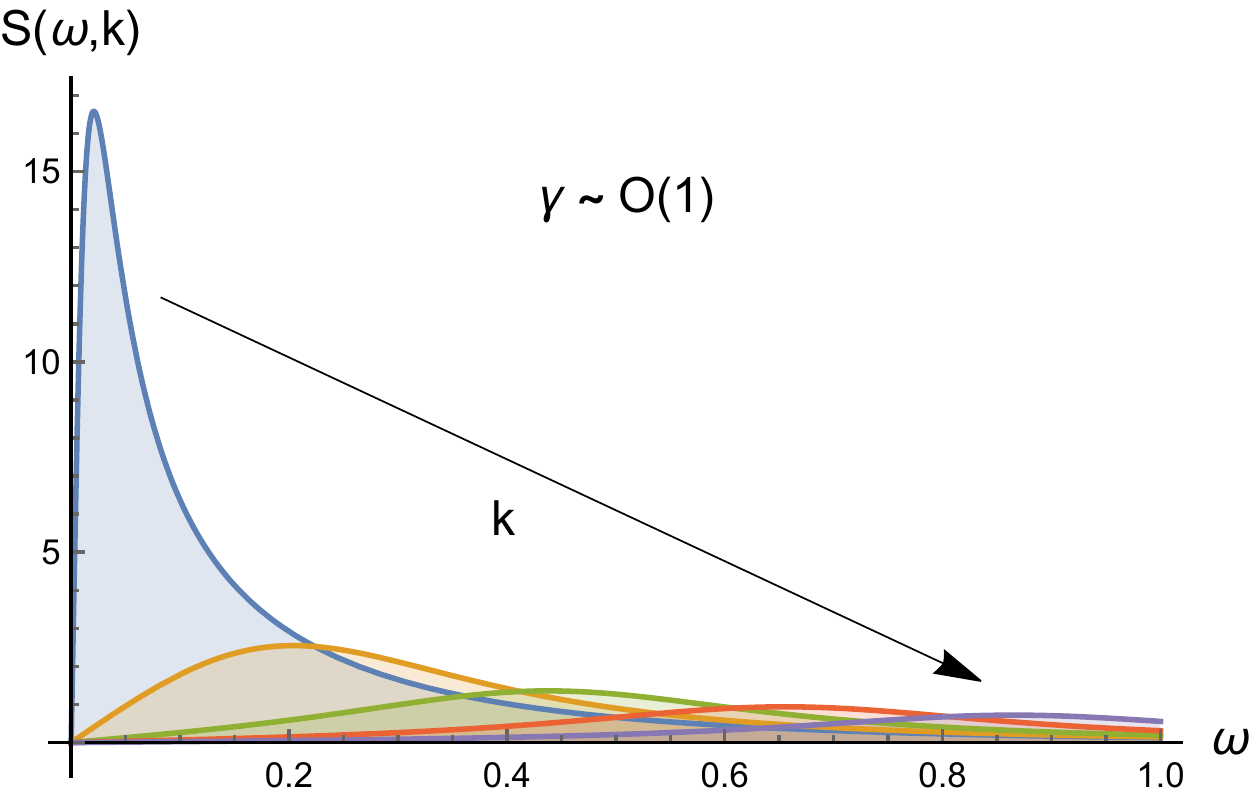}
    \caption{Phason spectral function, Eq.\eqref{sp}, for small and large damping $\gamma$. The phason speed is fixed to $v=1$. The top panel refers to $\gamma \approx 0$, while the bottom on to $\gamma=0.5$.}
    \label{fig2}
\end{figure}
From the Green's function in Eq.\eqref{gr}, it is straightforward to derive the corresponding spectral function which reads
\begin{equation}\label{sp}
   \mathcal{S}(\omega,k) \equiv -\frac{1}{\pi}\mathrm{Im}  \mathcal{G}(\omega+i \delta,k)=\frac{1}{\pi}\,\frac{\gamma \omega}{\left(\omega^2-v^2k^2\right)^2+\gamma^2 \omega^2}
\end{equation}
and is shown in Fig.\ref{fig2} for two different values of the phason damping. For small $\gamma$, in the underdamped regime, we see the typical feature of a propagating sound wave. On the contrary, for $\gamma \sim \mathcal{O}(1)$, we observe an incoherent response characterized by a flat and smoothed out spectrum. Notice that at this point the spectral function in Eq.\eqref{sp} is not normalized. For the discussion of the density of states and the heat capacity this is irrelevant, since we will be only interested in the low-energy scaling. Nevertheless, for the study of superconductivity, that is important. Therefore, we will later normalize it to compute the Eliashberg function.
\section{Density of states}
From the spectral function, Eq.\eqref{sp}, we can derive the phason density of states using 
\begin{equation}
    g(\omega)\equiv \frac{2\, \omega}{k_D^3}\int_0^{k_D}\, \mathcal{S}(\omega,k)\,k^2\,dk,
\end{equation}
where $k_D$ is taken to be the Debye cutoff, or in other words the maximum wave-vector allowed. Here, and in the rest of this manuscript, we only consider the case of a three dimensional system.

For small values of the damping, $\gamma \ll 1$, and at small frequency, we recover Debye law
\begin{equation}
    g(\omega)\approx \frac{\omega^2}{\omega_D^3}\equiv  g_D(\omega) \qquad \text{for}\qquad \gamma \ll 1,
\end{equation}
where $\omega_D= k_D v$ is the Debye frequency that serves as an ultraviolet cutoff for the phason frequency. This result is certainly not surprising since in the limit of small damping the phason is a sound-like propagating acoustic wave $\omega=\pm k$, as standard acoustic phonons. Therefore, in such a limit, it just plays the role of an extra acoustic wave obeying Debye's law.\\ 
When the damping becomes a non-negligible number, $\gamma \sim \mathcal{O}(1)$, the low-frequency scaling is modified to
\begin{equation}
    g(\omega)\approx \frac{\sqrt{\gamma } \,\omega ^{3/2}}{\sqrt{2} \,\omega_D^3} \qquad \text{for}\qquad \gamma \sim \mathcal{O}(1)\,,
\end{equation}
which can be re-written as:
\begin{equation}
    g(\omega)= \frac{\sqrt{\omega^\star}\omega^{3/2}}{\omega_D^{3}}\,.
\end{equation}
In other words, the Debye scaling is modified as $\omega^2 \rightarrow \sqrt{\omega^\star}\omega^{3/2}$. This also illustrates that the onset of this second regime is equivalent to the condition $\omega^\star>\omega$, which becomes more and more evident when $\gamma$ becomes large.
In the limit of very large $\gamma$, the $\omega^{3/2}$ regime is relegated to very low-frequencies, and a plateau emerges with value
\begin{equation}
    g(\omega)\approx \frac{2}{3\pi\,\gamma} \qquad \text{for}\qquad \gamma \gg 1\,.
\end{equation}
A constant density of states in three dimensions is exactly what one expects from a diffusive dynamics, \textit{cfr.} classical liquids, in which $g(0)$ is proportional to the diffusion constant \cite{hansen2013theory}. In the limit of large $\gamma$, the second damped mode $\omega=-i \gamma$ which appears from Eq.\eqref{gr}, is highly damped and therefore negligible. The only low-energy mode left is the diffusive phason, with diffusion constant $D_\gamma$, Eq.\eqref{diff}. Using the expression for the phason diffusion constant, the value of the plateau in the density of states can be rewritten as
\begin{equation}
    g(\omega)\sim \frac{D_\gamma}{v^2} \qquad \text{for}\qquad \gamma \gg 1\,,
\end{equation}
which is exactly of the same form as the diffusive component in liquids, where the diffusion constant there is related to self-diffusion \cite{hansen2013theory}.\\

\begin{figure}
    \centering
    \includegraphics[width=0.7\linewidth]{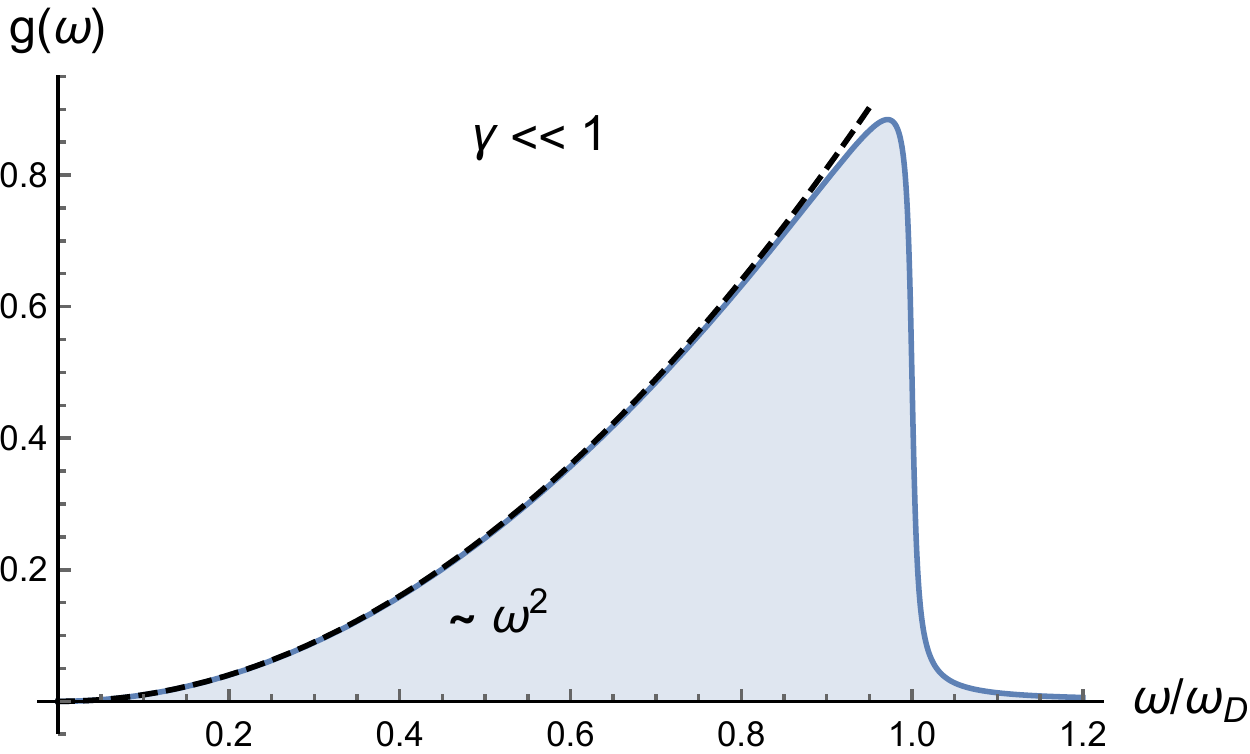}

    \vspace{0.1cm}
    
    \includegraphics[width=0.7\linewidth]{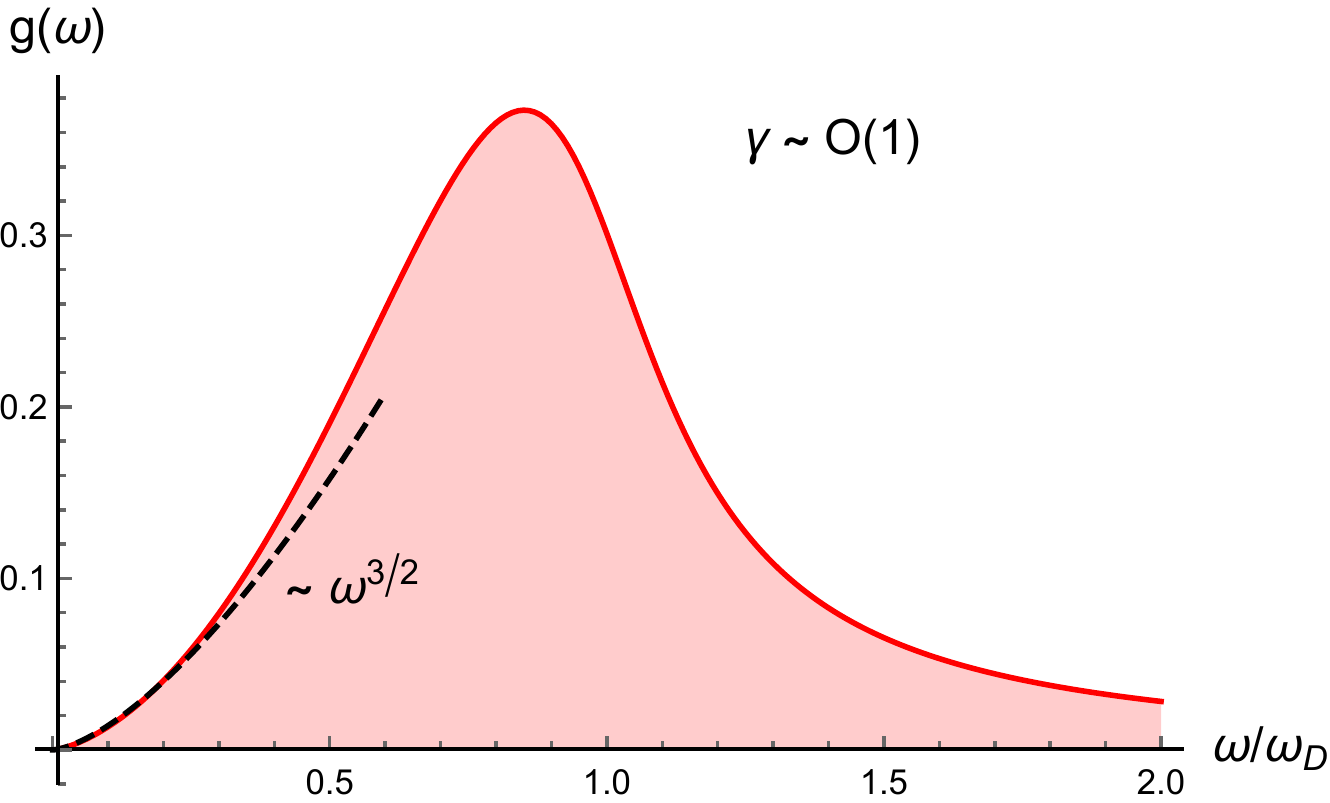}

      \vspace{0.1cm}

    \includegraphics[width=0.7\linewidth]{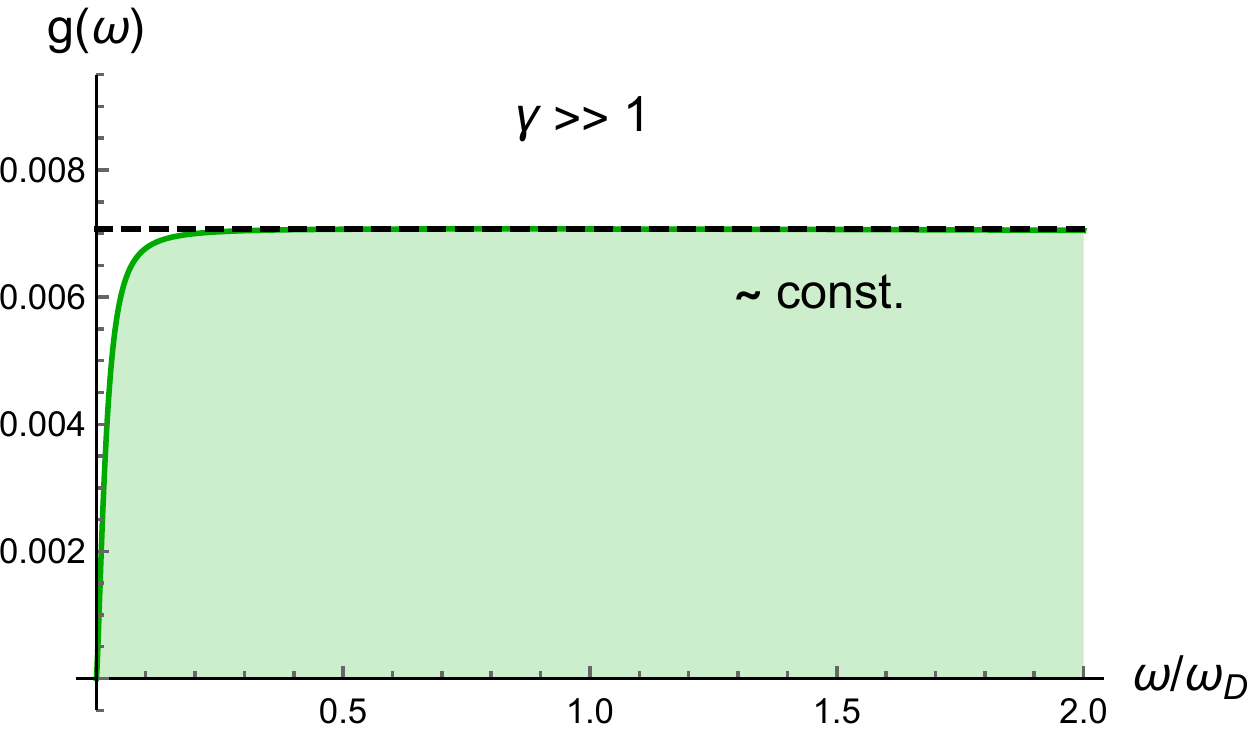}

    \caption{Phason density of states in the three different regimes $\gamma \ll 1$, $\gamma \sim \mathcal{O}(1)$ and $\gamma \gg 1$. For simplicity, we set $k_D=v=1$ and $\gamma=10^{-2},0.4,30$ from top to bottom.}
    \label{fig3}
\end{figure}

In summary, the phason density of states interpolates between the Debye expression for solids to the diffusive result for liquids by increasing the damping $\gamma$. All these three different regimes are shown in Fig.\ref{fig3}, where the dashed lines emphasize the different scalings discussed above.

Before continuing, a comment is in order. So far in the manuscript, we have been sloppy and indicated the different regimes using solely the damping parameter $\gamma$. This is not a totally correct practice since the latter is a dimensionful quantity. The correct comparison should be between the wave-vector scale at which the phason starts propagating, $\bar k$ or equivalently $k^\star$, and the maximum wave-vector allowed, $k_D$. In particular, the underdamped regime should correspond to the condition of $\gamma \ll v k_D$, in which the phason propagates at almost all scales. On the contrary, the overdamped condition should be given by $\gamma \gg v k_D$ and corresponds to no phason propagation at any scale. Since in the whole manuscript we have fixed $v=k_D=1$, the two conditions above correspond to what already used. In other words, whenever we present $\gamma$ as a scale, one should rather think of the dimensionless ratio $\gamma/v k_D$. In order to avoid clutter, we will not introduce an additional symbol for that dimensionless ratio.
\section{Heat capacity}
In order to proceed with the computation of the phason heat capacity, we assume a Bose-Einstein distribution for the phason modes and compute their contribution to the heat capacity by integrating the DOS
\begin{equation}
    C(T)=c_1\,k_B\,\int _0^\infty \,\left(\frac{\omega}{2k_B T}\right)^2\,\sinh \left(\frac{\omega}{2k_B T}\right)^{-2}\,g(\omega)d\omega,
\end{equation}
as if they were ``standard'' phonons.
Here, $c_1$ is a normalization factor which will not play any important role for our analysis and will be fixed to $1$ in the rest of the manuscript.\\
In the limit of small damping, we recover Debye law
\begin{equation}
   C(T)\propto T^3  \qquad \text{for}\qquad \gamma \ll 1,
\end{equation}
which is just a consequence of the Debye-like density of states of the underdamped phasons.
For non-negligible values of the damping, we obtain a slower than Debye scaling law
\begin{equation}
   C(T)\propto \sqrt{\gamma}~ T^{5/2} \qquad \text{for}\qquad \gamma \sim \mathcal{O}(1).
\end{equation}
In the limit of extremely large damping, the previous scaling is pushed to very small temperature, and  the heat capacity becomes approximately linear
\begin{equation}
   C(T)\propto \frac{T}{\gamma}\propto D_\gamma T  \qquad \text{for}\qquad \gamma \gg 1\,.
\end{equation}
Notice how the damping dependence here is different from the results of \cite{PhysRevLett.93.245902}. There, a finite pinning frequency has been assumed and the limit of $\omega_0 \rightarrow 0$ cannot be directly taken. It would be interesting to understand this difference more in detail.

\begin{figure}
    \centering
    \includegraphics[width=0.7\linewidth]{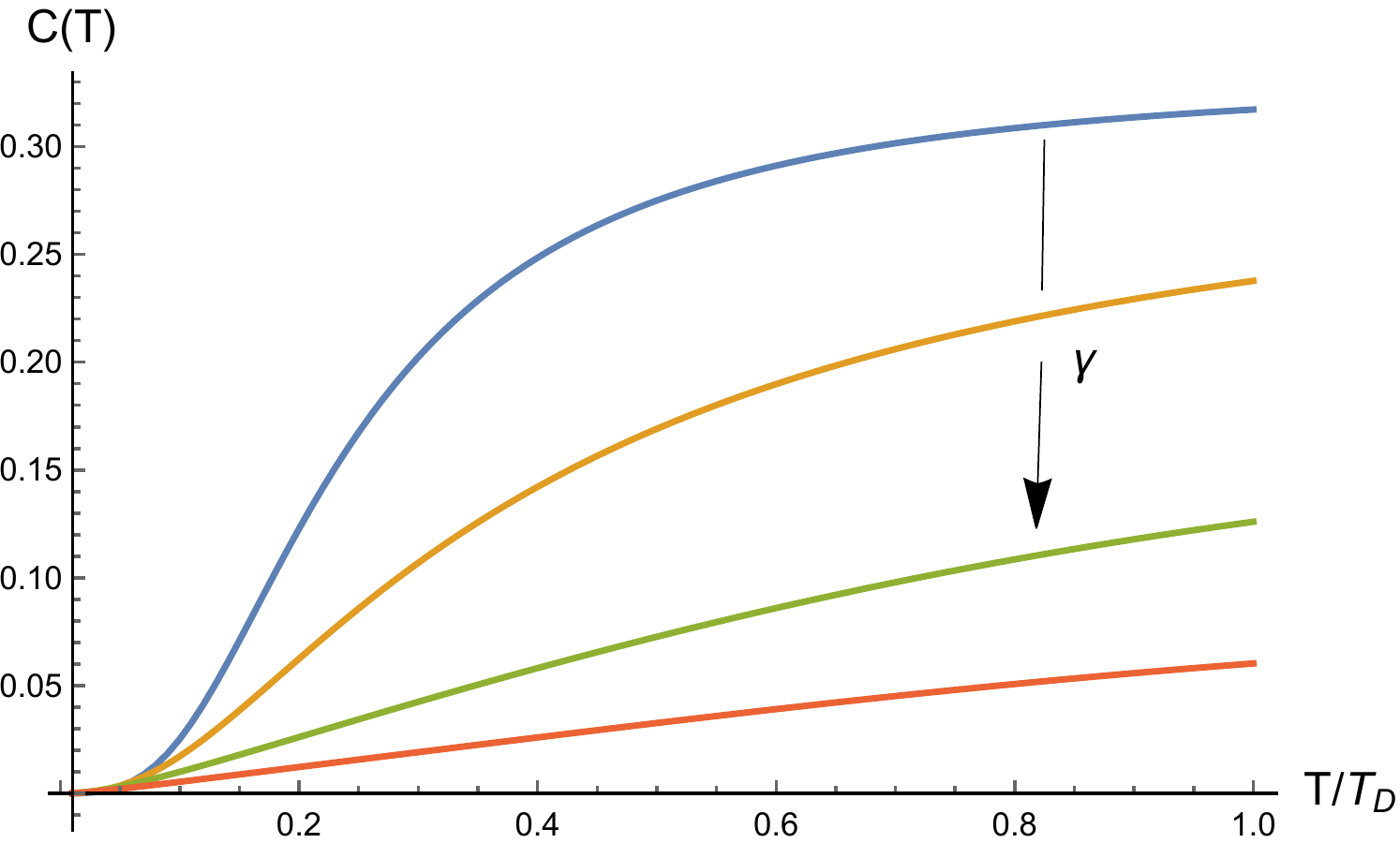}

    \caption{Phason heat capacity as a function of the reduced temperature increasing the phason damping. Here $v=1,k_D=1,k_B=1$ and $\gamma$ from $10^{-3}$ to $10$.}
    \label{fig4}
\end{figure}

\begin{figure}
    \centering
    \includegraphics[width=0.7\linewidth]{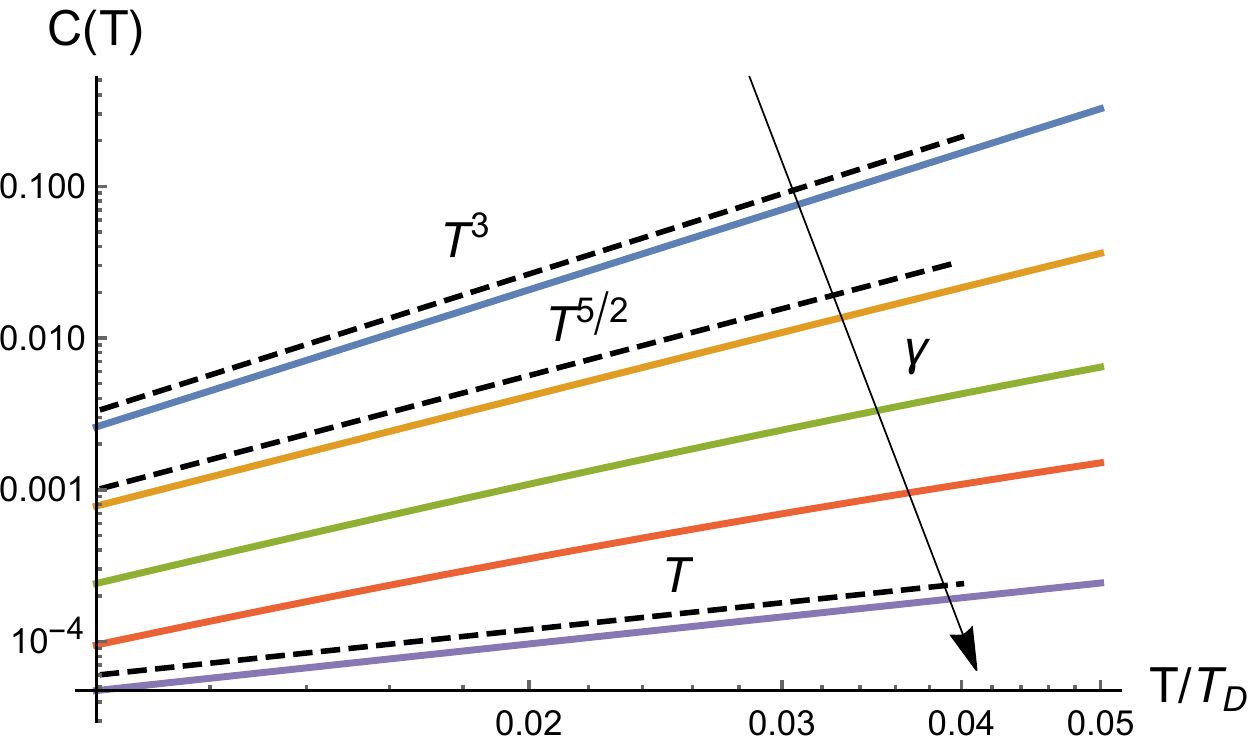}

    \caption{Low-temperature scaling of the phason heat capacity increasing phason damping. Here, we set $k_D=k_B=v=1$ and $\gamma$ from $10{-3}$ to $10^{3}$. Curves are shifted vertically for clarity.}
    \label{fig5}
\end{figure}

Continuing with our analysis, the general trend of the heat capacity is shown in Fig.\ref{fig4}.
The low-temperature scalings for different values of the phason damping are shown in Fig.\ref{fig5}, and they confirm the theoretical predictions.

Our results are compatible with the experimental data in \cite{PhysRevB.34.4432,PhysRevLett.76.2334}, where a quasi-linear scaling of the heat capacity is observed in incommensurate compounds. In the incommensurate dielectric (ClC$_6$H$_4$)$_2$S)$_2$ \cite{PhysRevLett.76.2334}, a scaling $C(T) \propto T^{1.7}$ is reported. One could probably claim such a number to appear as a combination of our $5/2$ and linear scalings. In the charge density compound charge-density-wave compound K$_{0.3}$MoO$_3$, a linear in $T$ scaling is observed, as in our theoretical model in the overdamped limit. Finally, let us repeat that similar observations regarding a linear in $T$ contribution from overdamped modes have already appeared in \cite{PhysRevLett.114.195502,PhysRevLett.93.245902,PhysRevLett.76.2334,doi:10.1142/S0217979221300024} using a slightly different formalism which includes a phason gap. Our results are in agreement with those presented there, where they overlap. Importantly, this indicates that overdamped bosonic modes would generally give a linear in $T$ contribution to the heat capacity, independently of the details of their dispersion.
\section{Superconductivity}
In this section, we imagine the existence of a possible pairing channel between electrons and phasons, and we try to estimate how the hypothetical critical temperature would depend on the phason parameters. For simplicity, we will only consider a s-wave pairing and treat the phasons as done in the case of phonons. 

\begin{figure}
    \centering
    \includegraphics[width=0.7\linewidth]{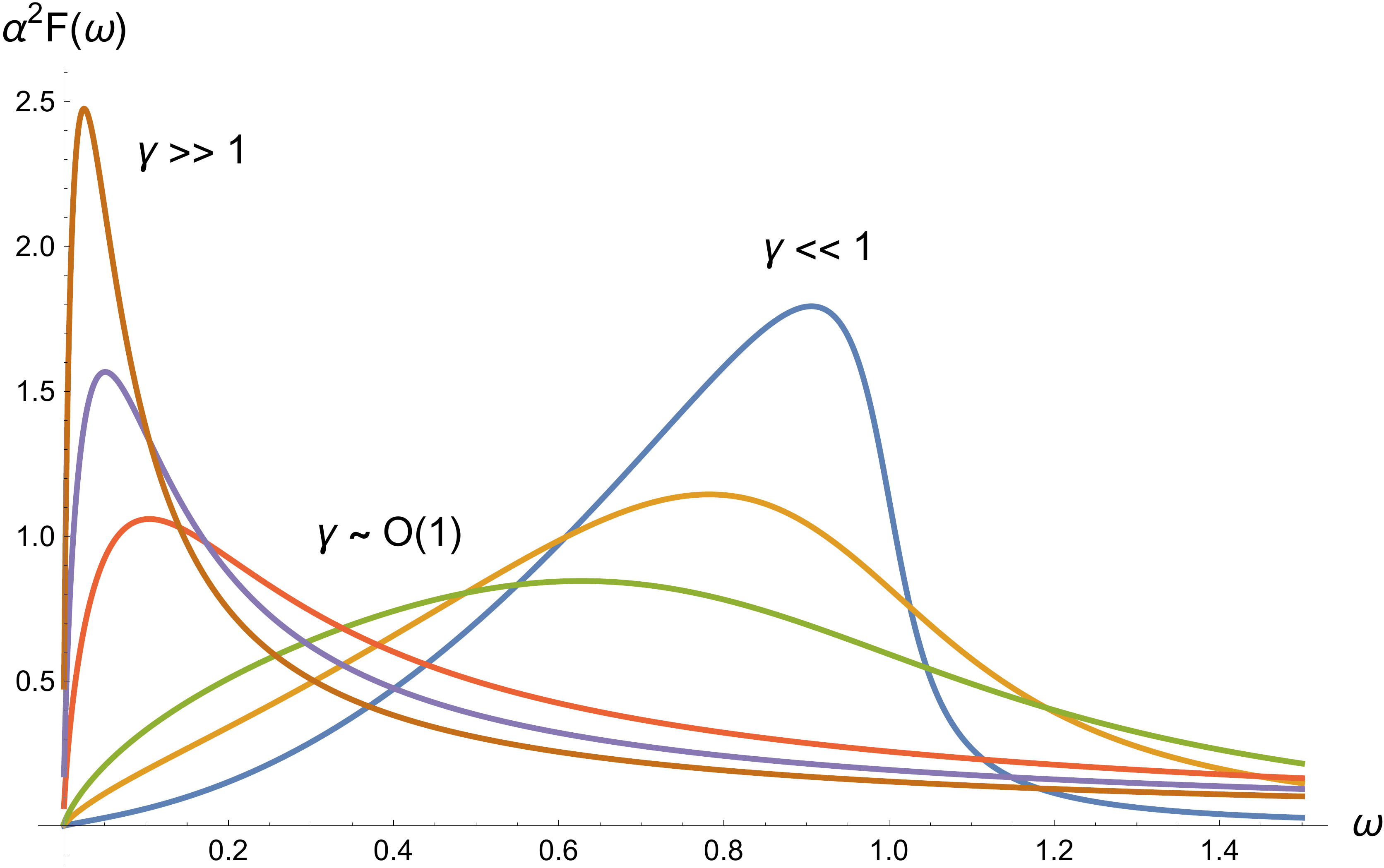}

    \caption{The Eliashberg function $\alpha^2 F(\omega)$ as a function of the frequency upon increasing phason damping. Here $v=k_D=1$, $c_2=0.3$ and $k_F=1/2$. $\gamma$ runs from $0.1$ to $50$.}
    \label{fig6}
\end{figure}

We start by defining the Eliashberg function
\begin{equation}
\alpha^2F(\boldsymbol{k},\boldsymbol{k'},\omega)\equiv \mathcal{N}(\mu)|g_{\boldsymbol{k},\boldsymbol{k'}}|^2 S(\omega,\boldsymbol{k}-\boldsymbol{k'}), \label{start point}
\end{equation}
where we assume $\mathcal{N}(\mu)$ constant and $g_q^2=C (vq)^2$. This second assumption, which is valid for acoustic phonons \cite{PhysRev.131.993,PhysRevB.3.3797}, is the simplest possibility and hence the one chosen. After a few algebraic manipulations and average over the Fermi surface, we can re-write the above expression as
\begin{equation}
    \alpha^2 F(\omega)=c_2\int_0^4 k_F^2\, \zeta\, S\left(\omega,k_F \sqrt{\zeta}\right) d\zeta, \label{acoustic result}
\end{equation}
where $c_2$ is a parameter to keep the normalized area $\int \alpha^2 F(\omega) d\omega=1$ constant. This is equivalent to normalize the phason spectral function as discussed in the previous sections.
The integral can be done analytically but the expression is rather cumbersome and therefore not shown. At low frequency, the Eliashberg function is linear in frequency:
\begin{equation}
    \alpha^2 F(\omega) \sim \gamma\, \omega +\dots
\end{equation}
Interestingly, this linear scaling is observed in amorphous superconductors \cite{keck1976superconductivity} (see also \cite{PhysRevB.3.3797}). We can prove that such a scaling is ubiquitous for overdamped bosonic modes with a constant in frequency/wavevector damping term, as $\gamma$.
The behavior of the normalized Eliashberg function is shown in Fig.\ref{fig6} for different values of $\gamma$.
The effective electron-phason coupling can be estimated as
\begin{equation}\label{lala}
\lambda=2 \int_{-\infty}^{\infty}\frac{ \alpha^2 F(\omega)}{\omega}d\omega\,.
\end{equation}
\begin{figure}
    \centering
    \includegraphics[width=0.7\linewidth]{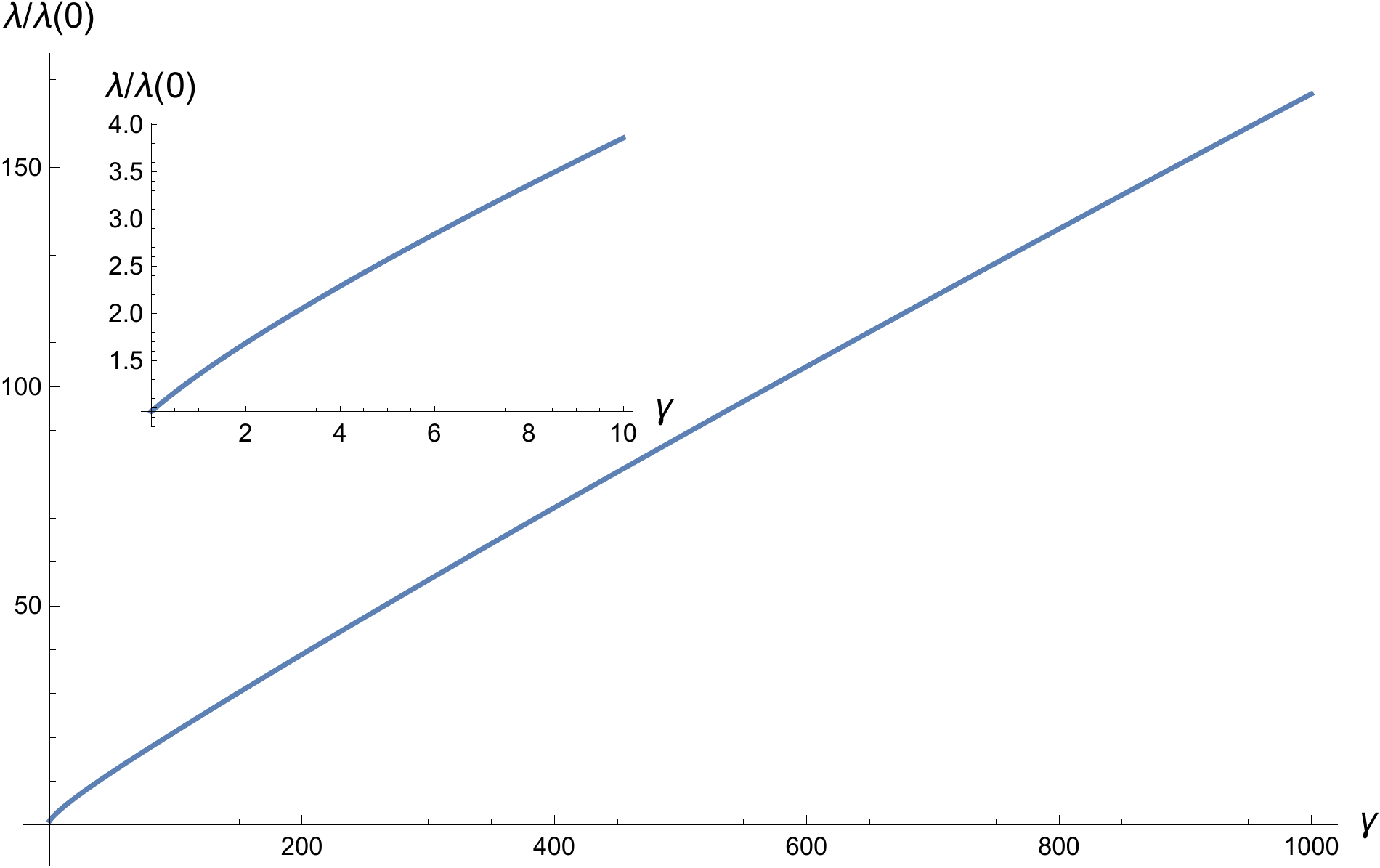}

    \caption{The effective electron-phason coupling $\lambda$ as a function of the phason damping normalized by its value at zero damping. Here $v=k_D=1$, $c_2=0.3$ and $k_F=1/2$. The inset zooms in the smaller damping region.}
    \label{fig7}
\end{figure}
Its dependence on the phason damping is shown in Fig.\ref{fig7}. It grows monotonically with the damping. This is just a consequence of the transfer of spectral weight to lower frequencies induced by the phason damping. This shows the possibility to achieve strongly-coupled superconductivity as an effect of incommensurability and the consequent phason modes, in agreement with the experimental findings of \cite{doi:10.1126/sciadv.aao4793}. Finally, using the simplified Allen-Dynes formula \cite{PhysRevB.12.905}, we can estimate the corresponding critical temperature
\begin{equation}
      T_c\,=\,\frac{\omega_{log}}{1.2}\,\exp\left(-\frac{1.04\,(1+\lambda)}{\lambda-u^\star\,-\,0.62\,\lambda\,u^\star}\right),\label{allenformula}
\end{equation}
where $\omega_{log}$ is the logarithmic average defined as:
\begin{equation}
    \omega_{log}=\exp\left[\int_0^\infty \frac{2}{\lambda \omega}\alpha^2F(\omega)\,\ln \omega\, d\omega\right].
\end{equation}
The latter is shown as a function of the phason damping in Fig.\ref{fig8}.
\begin{figure}
    \centering
    \includegraphics[width=0.7\linewidth]{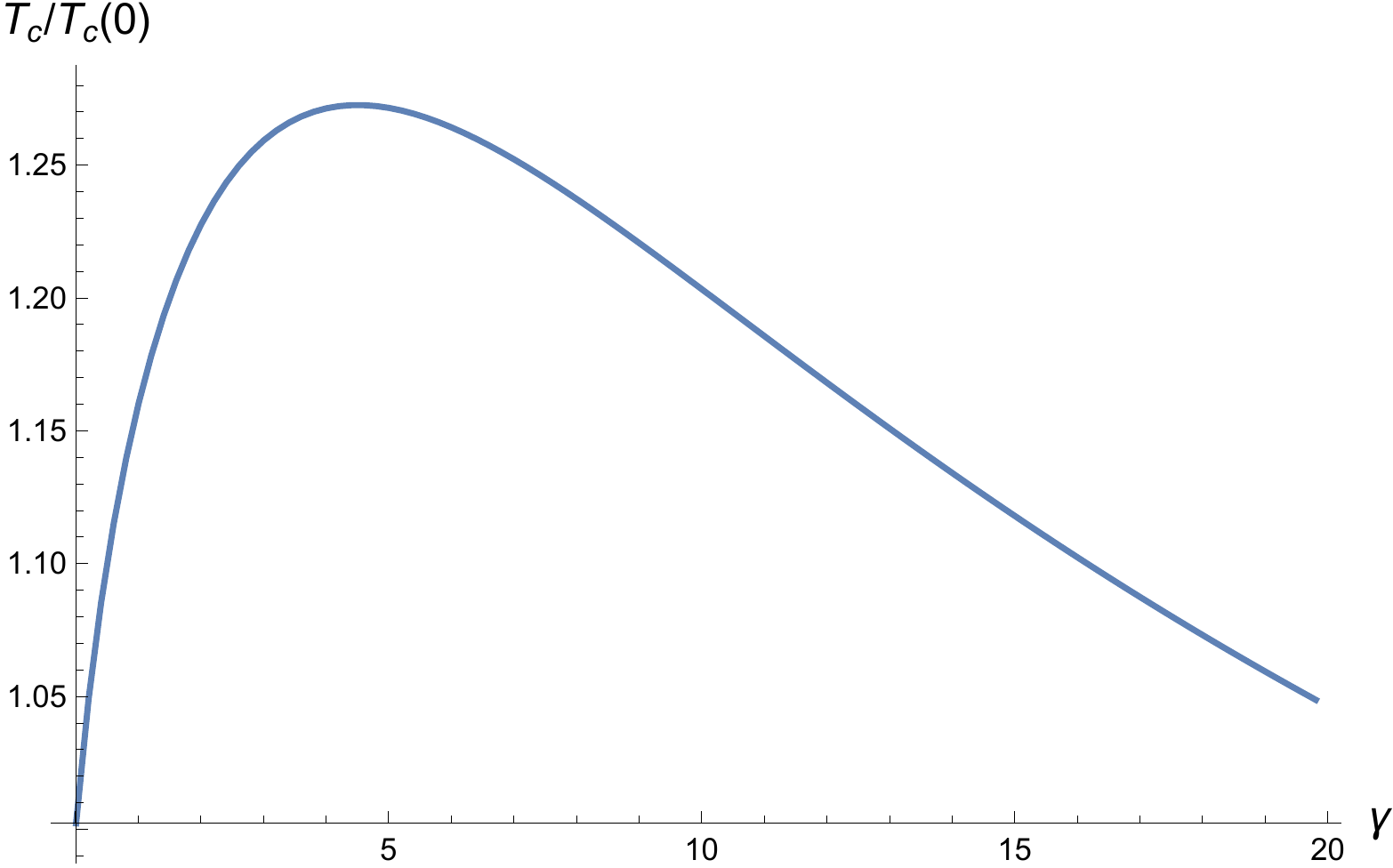}

    \caption{The superconducting critical temperature $T_c$ as a function of the phason damping. Here $v=k_D=1$, $c_2=0.3$ and $k_F=1/2$, $\mu^*=0.1$.}
    \label{fig8}
\end{figure}
Interestingly, for small damping, the critical temperature grows, up to a maximum critical value after which the trend is inverted. The reason for this non-monotonic is that while $\lambda$ always grows with damping, then, the average logarithmic frequency does not, and it decreases for large damping as shown in Fig.\ref{fig9}. Interestingly, we find that this change of behavior happens for $\gamma \approx 1$. This seems to suggest that in the underdamped regime, the phason damping enhances the critical temperature, while in the overdamped regime its effect is opposite. This behavior is similar to that observed for spin glass phase in cuprates~\cite{Setty2019} and optical modes in \cite{PhysRevB.102.174506,Jiang_2023}, and it is expected to persist also in the limit of finite pinning frequency, $\omega_0 \neq 0$.

Let us discuss our results in comparison to the existing literature. First, the strong enhancement of the effective coupling visible in Fig.\ref{fig8} is compatible with the experimental observations in \cite{doi:10.1126/sciadv.aao4793}, where the strongly coupled nature of superconductivity has been ascribed to the phason mode and its red-shifted spectral weight, as a consequence of its overdamped nature. Second, recent theoretical analysis \cite{oliveira2023incommensurability} has shown that incommensurability can provide an enhancement of the superconducting critical temperature, but only in the regime of weak or intermediate coupling. Our findings offer a different but compatible interpretation of the same fact. Phasons, and in particular phason damping, enhance the critical temperature $T_c$ only in the regime of small damping (\textit{i.e.}, the underdamped regime), in which the Eliashberg coupling is not too large (see Fig.\ref{fig8}), as advocated in \cite{oliveira2023incommensurability}.
\begin{figure}
    \centering
    \includegraphics[width=0.7\linewidth]{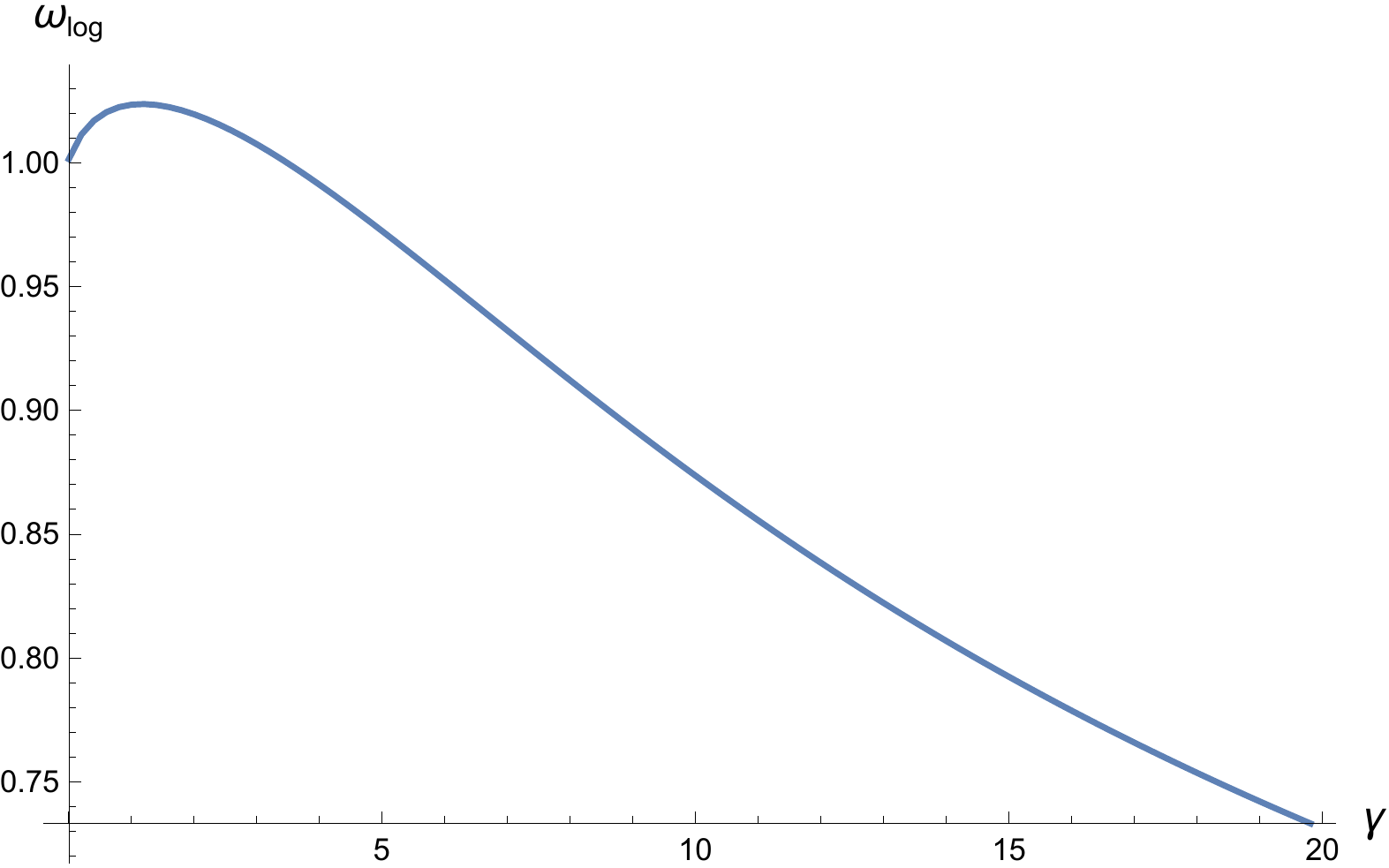}

    \caption{The average $\omega_{log}$ as a function of the phason damping. Here $v=k_D=1$, $c_2=0.3$ and $k_F=1/2$.}
    \label{fig9}
\end{figure}
\section{Discussion}
In this work, we have studied in detail the effects of phason modes on the vibrational, thermodynamic and superconducting properties of incommensurate structures. For simplicity, we have worked in the limit in which the phason pinning frequency is negligible compared to its damping. Our analysis revealed several interesting features which appear, at least qualitatively, compatible with the existing experimental data and other theoretical studies in the literature.

We have confirmed that overdamped phasons, or more in general overdamped excitations, induce glassy-like features in the heat capacity, and in particular a pronounced linear in $T$ contribution, which is proportional to their diffusive constant and more evident in the limit of very large damping. We have also found an intermediate scaling regime, which combined with the linear scaling emerging at large damping, could explain the quasi-linear behaviors observed experimentally in certain incommensurate compounds \cite{PhysRevB.34.4432,PhysRevLett.76.2334}. Interestingly, our results confirm the previous theoretical studies in \cite{PhysRevB.99.054305,PhysRevLett.93.245902,PhysRevLett.114.195502}, but also suggest a different scaling of the linear in $T$ contribution with the phason damping. The reason of the discrepancy is due to the different limits taken. In particular, we assume from the start that the pinning frequency is negligible, while the works referred above do not. It would be interesting to investigate further this point.

Moreover, inspired by the results in \cite{ochoa2023extended,PhysRevB.100.155426}, we have investigated a hypothetical phason-induced superconductivity using the Eliashberg formalism, and treating naively the phasons as if they were standard phonons. Despite the simplicity of our assumptions, which certainly necessitate further validation in the near future, we have been able to extract several interesting features which appear in agreement with recent observations. First, we have shown that, because of the overdamped nature of phasons (which is usually not the case for phonons, unless in situations with very strong anharmonicity), the effective coupling can be strongly enhanced, leading to strongly-coupled superconductivity. Similar results, and experimental verifications, have been presented in \cite{doi:10.1126/sciadv.aao4793} for the quasiperiodic host-guest structure of elemental bismuth at high pressure, Bi-III. Additionally, we have predicted a non-monotonic behavior of the superconducting critical temperature $T_c$ as a function of the phason damping. The maximum in $T_c$ appears approximately at the underdamped to overdamped crossover scale, indicating that only underdamped phason contribute positively to superconductivity. Qualitatively, this finding is in agreement with the recent results in \cite{oliveira2023incommensurability}, since the underdamped regime corresponds to the range of weak or intermediate effective coupling $\lambda$. This observation deserves further studies.

We foresee several improvements of our setup and arguments against its simplicity. As a concrete example, we have been completely agnostic about the microscopic details of a possible phason-electron coupling, and even about the possibility that the latter exists. Preliminary studies of the phason-electron coupling in twisted bilayer graphene have appeared in \cite{PhysRevB.100.155426,PhysRevLett.128.065901,ochoa2023extended}, and suggested that at least for such a scenario the pairing mechanism might be more complicated than what we have modelled (\textit{e.g.}, dependent on the twist angle $\theta$, etc.). Additionally, it is not clear to us up to which point phasons could be treated exactly as standard phonons. In our work, we have blindly assumed that is the case.

Despite the simplicity of our analysis, and the bold assumptions, we believe that this might constitute a first baby step towards a deeper understanding of phason properties and their effects on thermodynamics, transport and superconductivity, which are potentially relevant for a large class of quantum materials, joined under the umbrella of incommensurate structures. We expect more and more signatures of these overdamped modes to appear in the near future. It would be important to think about possible measurable effects arising from the presence of an overdamped phason mode. The analysis of the diffusive dynamics of charge order might be a good place to start \cite{PhysRevB.100.205125,doi:10.1126/sciadv.aax3346}.
\section*{Aknowledgments}
We thank Hector Ochoa for useful comments and suggestions on a preliminary version of this manuscript. We would like to thank Friedrich Malte Grosche for telling us about Ref.\cite{Pippard_1987} and for pointing out to us the existence of 1D-like phason modes in certain compounds. We would like to thank Matteo Mitrano for discussions about the possible experimental consequences of an overdamped phason. We thank S.~Nakamura, J.~Douglas, Y.~Wang, M.~Landry for several discussions about phasons and their existence. We thank G. ~Ghiringhelli, R.~Arpaia, M.~Grosche, S.~Mandyam, Y.~Ishii and J.~Ma for useful discussions about superconductivity and incommensurate structures.
C.J. and M.B. acknowledge the support of the Shanghai Municipal Science and Technology Major Project (Grant No.2019SHZDZX01) and the sponsorship from the Yangyang Development Fund. A.Z. gratefully acknowledges funding from the European Union through Horizon Europe ERC Grant number: 101043968 ``Multimech'', and from US Army Research Office through contract nr.W911NF-22-2-0256.

\appendix

\section{Comparison to the heat capacity of a damped harmonic oscillator}
\begin{figure}[h]
    \centering
    \includegraphics[width=0.9\linewidth]{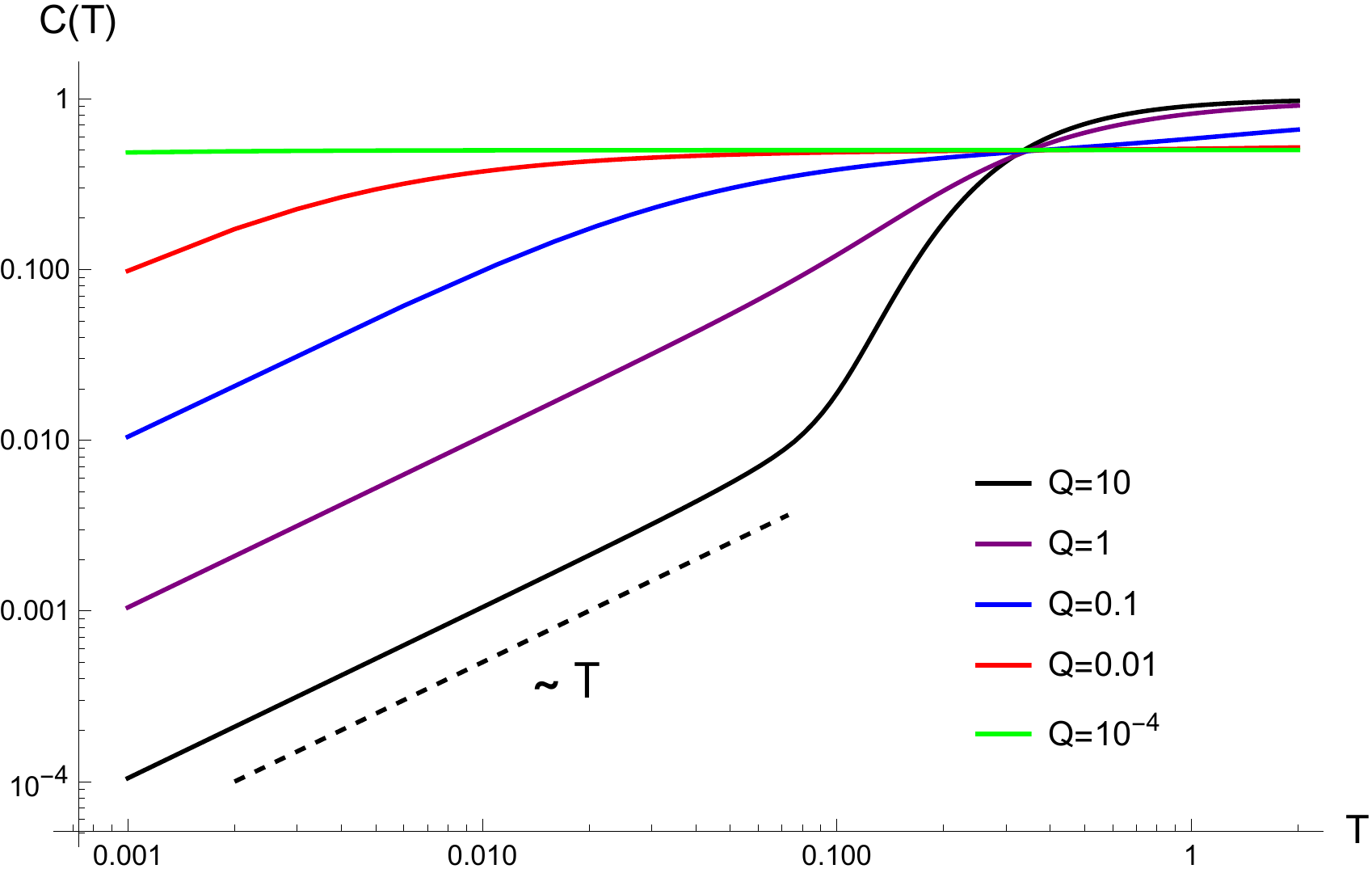}
    \caption{The heat capacity of a damped harmonic oscillator obtained from Pippard's analysis \cite{Pippard_1987} for different values of the parameter $Q$. For simplicity here $k_b=\bar \omega=1$. The dashed line indicates the linear in $T$ scaling.}
    \label{fig:pipp}
\end{figure}

In 1987, Pippard computed the entropy of a damped harmonic oscillator using an analogy with the physics of RLC circuits \cite{Pippard_1987}. The final result of such an analysis, based on classical physics, is that the entropy of a damped harmonic oscillator is given by:
\begin{equation}
    S=\frac{k_B Q}{\pi} \int_0^\infty \frac{(x^2+1)\left[\beta x \coth (\beta x)-\ln (2 \sinh (\beta x))\right]}{x^2+Q^2 (x^2-1)^2}dx
\end{equation}
where $\beta=\bar \omega/2 k_B T$ (with $\bar \omega$ the characteristic frequency of the harmonic oscillator), and $Q$ is the quality factor $Q=\bar \omega/\gamma$ where $\gamma$ is the damping parameter. In particular, by decreasing the value of $Q$ the harmonic oscillator passes from being underdamped to overdamped. Using the above expression for the entropy, one can compute the heat capacity using:
\begin{equation}
    C(T)=T\frac{\partial S}{\partial T}\,.
\end{equation}
The results of this simple computation are shown in Fig.\ref{fig:pipp} where the heat capacity as a function of temperature is plotted for different values of the parameter $Q$. Interestingly, we observe a clear linear in $T$ regime at low temperature. The range in which this scaling appears becomes smaller by decreasing $Q$. This is compatible, at least qualitatively, with our results, in which a overdamped phason also gives rise to a linear in $T$ contribution to the heat capacity. It would be interesting to explore further the relation between our results and Pippard's computations \cite{Pippard_1987}. Notice that our computation relies on the quantum bosonic nature of the phason excitations at low temperature, while Pippard's analysis is purely classical. Interestingly, a linear in $T$ heat capacity has been obtained as well for a damped quantum oscillator in \cite{PhysRevE.79.061105}. This strongly hints towards the universality of this behavior in the context of overdamped modes.
\section{Overdamped phason with restricted mobility}
In high-pressure bismuth \cite{doi:10.1126/sciadv.aao4793} and other host-guest structures, the phason dispersion appears to be 1D-like with a strong dispersion along the chains, but flat dispersion perpendicular to the chains. Therefore, in this appendix, we find interesting to repeat our computation by considering an overdamped phason mode which propagates only in some of the three spatial directions. This amounts to modifying the equation for the density of states as:
\begin{equation}
    g(\omega)\propto\, \int_0^{k_D} \frac{\gamma \omega}{\left(\omega^2-\omega_0^2-v^2k^2\right)^2+\gamma^2 \omega^2}\,k^n\,dk,
\end{equation}
where $n=0,1,2$ correspond respectively to 1D, 2D and 3D propagating phasons, and we have added a bare energy $\omega_0$ to take into account the flat dispersion in the other directions. By analytical expanding the above expression in the overdamped regime, defined as $\gamma$ much larger than every other frequency scale, we obtain that the VDOS at low frequency is constant, and indeed proportional to $1/\gamma$, independently of $n$. That is to say that, independently of the 1D,2D or 3D nature of the overdamped phason, in such a regime, that will always contribute to the heat capacity with a linear in $T$ term, which is in principle observable in experiments. Let us also emphasize that in the overdamped limit $\gamma \gg \omega_0$ and therefore an optical-like and a propagating phason would give exactly the same result. This is compatible with the results obtained in the main text and the comparison with those of \cite{PhysRevLett.93.245902}, where a bare energy (or pinning frequency) $
\omega_0$ for the phason is considered. The main difference in the presence of a finite $\omega_0$ is the appearance of a boson-peak-like feature at larger energies (see \cite{PhysRevLett.93.245902} for details). If we insist on ignoring the energy gap $\omega_0$, and we set it to zero from the start, then for intermediate values of the damping coefficient we find that $g(\omega)\propto \omega^{(n+1)/2}$ and $C(T)\propto T^{(n+1)/2+1}$ where $n+1$ is the number of spatial directions along which the phason propagates.
\bibliographystyle{apsrev4-1}
\bibliography{ref}
\end{document}